%% file: notes_mfo-mso.tex
\documentclass[runningheads, envcountsame, a4paper]{llncs}
\usepackage[english]{babel}
\usepackage{amsmath}
\usepackage{amssymb}
\usepackage{colortbl}
\usepackage{txfonts}
\usepackage{xypic}
\usepackage{multirow}
\usepackage{textcomp}
\usepackage{calc}
\usepackage{stmaryrd}
\usepackage{marvosym}
\usepackage{pgf}
\usepackage{ifthen}
\usepackage{mdwlist}
\usepackage{enumerate}
\usepackage{paralist}
\usepackage{xcolor}
\usepackage{bm}
\usepackage{cancel}
\usepackage{pdfpages}
\usepackage[hidelinks]{hyperref}
\usepackage{todonotes}

\usepackage{tikz}
\usetikzlibrary{shapes.geometric}
\usetikzlibrary{arrows,automata}
\usetikzlibrary{matrix}
\usetikzlibrary{decorations.pathmorphing}
\usetikzlibrary{decorations.markings}
\usetikzlibrary{decorations.shapes}

\usepackage{caption}
\usepackage{subcaption}
\usepackage{url}
\usepackage{float}
\floatstyle{ruled}
\newfloat{program}{thp}{lop}
\floatname{program}{Algorithm}
\newfloat{program-schema}{thp}{lop}
\floatname{program-schema}{Algorithm schema}

\newcommand{\powerset}{\mathcal{P}}

\DeclareMathOperator{\sux}{succ}
\DeclareMathOperator{\Sux}{Succ}
\DeclareMathOperator{\Sing}{Sing}

\newcommand{\suchthat}{\;\ifnum\currentgrouptype=16 \middle\fi|\;}

\newcommand{\mrk}[1]{{#1}'}
\newcommand{\oldstack}[3]{%
{\ifthenelse{\equal{#1}{1}}{%
\mrk{#2}
}%
{#2}}_{#3}%
}
\newcommand{\stack}[3]{%
[%
{\ifthenelse{\equal{#1}{1}}{%
\mrk{#2}
}%
{#2}}\ {#3}%
]%
}

\makeatletter
\newcommand{\shift}[2][\hspace*{4mm}]{\ext@arrow 0359\rightarrowfill@@{#1}{#2}}
\def\rightarrowfill@@{\arrowfill@@\relax\relbar\rightarrow}
\def\arrowfill@@#1#2#3#4{%
  $\m@th\thickmuskip0mu\medmuskip\thickmuskip\thinmuskip\thickmuskip
   \relax#4#1
   \xleaders\hbox{$#4#2$}\hfill
   #3$%
}
\makeatother

\newcommand{\g}[1]{\textbf{\textit{#1}}}

\newcommand{\deffmla}{\ensuremath{\stackrel{\text{def}}{=}}}
\DeclareMathOperator{\last}{last}
\DeclareMathOperator{\first}{first}

\newtheorem{statement}{Statement}

\newcommand{\tagxml}[2]{%
\langle %
{\ifthenelse{\equal{#1}{1}}{%
/#2  
}%
{#2}}%
\rangle%
}

\newif\ifemptystrok
\emptystroktrue

\makeatother

\makeatletter
\renewcommand*{\@fnsymbol}[1]{\ensuremath{\ifcase#1\or \ifemptystrok \in \else \notin \fi\or \dagger\or \ddagger\or
    \mathsection\or \mathparagraph\or \|\or **\or \dagger\dagger
    \or \ddagger\ddagger \else\@ctrerr\fi}}
\makeatother

\tikzset{
	path/.style={dotted},
	every edge/.style={draw,solid},
	normal/.style={solid},
	siblings/.style={dashed},
	support/.style={decorate, decoration={snake,amplitude=.4mm,segment length=2mm,post length=1mm}}
}

\begin{document}
\title{Lecture Notes on\\Monadic First- and Second-Order Logic on Strings}
	\author{
        Dino Mandrioli \and Davide Martinenghi \and Angelo Morzenti \and\\ Matteo Pradella  \and Matteo Rossi
		}
	\institute{
			Dipartimento di Elettronica, Informazione e Bioingegneria (DEIB), 
            Politecnico di Milano, Piazza Leonardo Da Vinci 32, 20133 Milano, Italy\\
		\email{\{firstname.lastname\}@polimi.it}
	}
\authorrunning{D. Mandrioli et al.}
\maketitle


\input{intro}
\input{mfo}

\input{mso}

\bibliography{opbib}

\end{document}

%% file: intro.tex
\section{Introduction}
\label{sec:into}

From the very beginning of formal language and automata theory, the investigation
of the relations between defining a language through some kind of abstract
machine and through a logic formalism has produced challenging theoretical
problems and important applications in system design and verification. A
well-known example of such an application is the classical Hoare's method to
prove the correctness of a Pascal-like program w.r.t. a specification stated as
a pair of pre- and post-conditions expressed through a first-order theory \cite{DBLP:journals/cacm/Hoare69}.

Such a verification problem is undecidable if the involved formalisms have the
computational power of Turing machines but may become decidable for less
powerful formalisms as in the important case of Regular Languages. Originally, B\"uchi, Elgot, 
and  Trakhtenbrot \cite{bib:Buchi1960a,Elg61,Tra61} independently developed a \emph{Monadic Second-Order
logic} defining exactly the Regular Language family, so that the decidability
properties of this class of languages could be exploited to achieve automatic
verification. 

Intuitively, monadic logics have some syntactic restrictions on the predicates used.
In particular, only predicates that have one argument (i.e., whose arity is 1) are allowed (with the exception of the ordering relation).
As usual, in the first-order case only variables can be quantified.
In the second-order, instead, monadic predicates---i.e., predicates with arity 1, as mentioned above---can also be quantified, thus resulting in so-called second-order variables.

Interestingly, to capture the full class of Regular Languages by means of a monadic logic it is necessary to exploit a second-order version of the logic, which is more powerful than the simpler first-order one; it has been shown by McNaughton and Papert \cite{McNauPap}, however, that restricting the logic to first-order allows users to define precisely the \emph{non-counting} subclass of Regular Languages---i.e., the languages which cannot ``count'' the number of repeated occurrences  of a given subword.\footnote{For instance, the language $\{(ab)^{2n} \mid n > 0\}$  is counting. Non-counting languages, in turn, are equivalent to other interesting subclasses of regular ones, such as, e.g., the \emph{star-free} ones, i.e. those languages that can be defined by means of regular expressions not making use of the Kleene-* operation.}

Such logic characterizations, however, have not been exploited in practice to achieve automatic verification due to the intractable complexity of the necessary algorithms. 
Later on, a major breakthrough in this field has been
obtained thanks to the advent of \emph{model checking}, which exploits language characterization in terms of \emph{temporal logic} \cite{Emerson90}. Temporal logic has the same expressive power as first-order logic but, being less succint, allows for more efficient (though still exponential) verification algorithms.

These notes present the essentials of first- and second-order  monadic logics with introductory purpose and are organized as follows. In Section~\ref{sec:MFO}, we discuss Monadic First-Order logic and show that it is strictly less expressive than Finite-State Automata, in that it only captures a strict subset of Regular Languages---the non-counting ones. We then introduce Monadic Second-Order logic in Section~\ref{sec:MSO}; such a logic is, syntactically, a superset of Monadic First-Order logic and captures Regular Languages exactly.
We also show how to transform an automaton into a corresponding formula and vice versa.
Finally, in Section~\ref{sec:discussion} we discuss the use of logical characterizations of classes of languages, such as those described in Sections~\ref{sec:MFO} and~\ref{sec:MSO}, as the basis for automatic verification techniques.

%% file: mfo.tex
\section{Monadic First-order Logic of Order on Strings}
\label{sec:MFO}


Given an input alphabet $\Sigma$, formulae of the \emph{monadic first-order logic} (MFO) are built out of the following elements:

\begin{itemize} 

\item
First-order variables, denoted as lowercase letters (written in boldface to avoid confusion with strings), $\bm{x}$, $\bm{y}$, \ldots, which are interpreted over the natural numbers $\mathbb N$.

\item
Monadic predicates $a(\cdot)$, $b(\cdot)$, \ldots, one for each symbol of $\Sigma$; intuitively, $a(\bm x)$ evaluates to true in a string $w$ if, and only if, the character of $w$ at position $\bm x$ is $a$.

\item
The order relation $<$ between natural numbers.

\item
The usual propositional connectives and first-order quantifiers.

\end{itemize}

More precisely, let $\mathcal{V}$ be a finite set of first-order variables, and let $\Sigma$ be an alphabet.
Well-formed formulae of the MFO logic are defined according to the following syntax:
\[
\varphi :=
           a(\bm x)
             \; \mid \;
           \g{x} < \g{y} 
             \; \mid \;
           \neg \varphi
             \; \mid \;
           \varphi \lor \varphi 
             \; \mid \;
           \exists \bm x (\varphi)
\]
where $a \in \Sigma $ and $\g{x}, \g{y} \in \mathcal{V}$.

    The usual
        predefined abbreviations are introduced to denote the remaining
        propositional connectives, the
        universal
        quantifier, the arithmetic relations $\geq, \leq, =, \ne, >$, and sums and subtractions between first order variables and numeric constants.

We have the following definitions of propositional connectives and first-order quantifiers:
$$
\begin{aligned}
\varphi_1 \land \varphi_2             \; \deffmla & \; \ \neg (\neg \varphi_1 \lor \neg \varphi_2) \\
\varphi_1 \Rightarrow \varphi_2      \; \deffmla & \; \ \neg \varphi_1 \lor \varphi_2   \\
\varphi_1 \Leftrightarrow \varphi_2  \; \deffmla & \; \ (\varphi_1 \Rightarrow \varphi_2) \land (\varphi_2 \Rightarrow \varphi_1)  \\
\forall \g{x} (\varphi)              \; \deffmla & \; \ \neg \exists \g{x} (\neg \varphi) \\
\end{aligned}
$$
the following definitions of relations:
$$
\begin{aligned}
\g{x} \geq \g{y}           \; \deffmla & \; \ \neg (\g{x} < \g{y}) \\
\g{x} \leq \g{y}           \; \deffmla & \; \ \g{y} \geq \g{x} \\
\g{x} = \g{y}              \; \deffmla & \; \ \g{x} \leq \g{y} \land \g{y} \leq \g{x} \\
\g{x} \neq \g{y}           \; \deffmla & \; \ \neg (\g{x} = \g{y})  \\
\g{x} > \g{y}              \; \deffmla & \; \ \g{y} < \g{x}  \\
\end{aligned}
$$
and the following definitions of constants, of the successor of a natural number, and of addition and subtraction of constant values:
$$
\begin{aligned}
\g{x} = 0               \; \deffmla & \; \ \forall \g{y} \neg (\g{y} < \g{x}) \\
\sux(\g{x}, \g{y})      \; \deffmla & \; \ \g{x} < \g{y} \land \neg \exists \g{z} (\g{x} < \g{z} \land \g{z} < \g{y})    \\
\g{y} = \g{x} + k       \; \deffmla & \; \ \exists \g{z}_0, \ldots, \g{z}_{k} (\g{z}_0 = \g{x} \land \sux(\g{z}_0,\g{z}_1) \land \sux(\g{z}_1,\g{z}_2) \land \ldots \land \sux(\g{z}_{k-1},\g{z}_k) \land \g{y} = \g{z}_{k}) \\
\g{y} = \g{x} - k       \; \deffmla & \; \ \g{x} = \g{y} + k
\end{aligned}
$$
where $k$ is a constant in $\mathbb N$. Further useful abbreviations are the follwowing ones:
\begin{itemize}
\item
$\first(\g{x})$ and $\last(\g{x})$ identify, respectively, the first and last positions in the string:
$
\first(\g{x}) \deffmla \neg \exists \g{y} (\g{y} < \g{x})
$, obviously equivalent to \g{x} = 0; \\
$
\last(\g{x}) \deffmla \neg \exists \g{y} (\g{y} > \g{x})
$
\end{itemize}

\ifemptystrok
An MFO formula is interpreted over a string $w \in \Sigma^*$,
with respect to assignment $\nu: \mathcal{V} \to U$, where $U = \{0, \ldots, |w|-1\}$, which maps $\mathcal{V}$ to a position in string $w$.  
Notice that, if $w=\varepsilon$, then function $\nu(\g{x})$ is undefined for every variable $\g{x} \in \mathcal{V}$.  
The satisfaction relation (indicated, as usual, as $\models$) for MFO formulae is defined in the following way:
\begin{itemize}
    \item
    $w, \nu \models a(\g{x})$ if, and only if, $\nu(\g{x})$ is defined and $w[\nu(\g{x})]=a$ 
    \item
    $w, \nu \models \g{x} < \g{y}$ if, and only if, $\nu(\g{x}) < \nu(\g{y})$ holds 
    \item
    $w, \nu \models \neg\varphi$ if, and only if, $w, \nu \not\models \varphi$ holds 
    \item
    $w, \nu \models \varphi_1 \lor \varphi_2$ if, and only if, at least one of $w, \nu \models \varphi_1$ and  $w, \nu \models \varphi_2$ holds 
    \item
    $w, \nu \models \exists \g{x}(\varphi)$ if, and only if, $|w|>0$ and $w, \nu[i / \g{x}] \models \varphi$ hold for some $i \in \{0,\,\ldots,\,|w|-1 \}$
\end{itemize}
where $w[k]$ denotes the letter of $w$ at position $k$, and $\nu[i / \g{x}]$ is the mapping that assigns $i$ to $\g{x}$ and otherwise coincides with $\nu$.
Notice that, in case $w=\varepsilon$ (and therefore $U=\emptyset$),  
$w, \nu \not\models a(\g{x})$ for any $a$ and $\g{x}$ (i.e., it is not the case that $w, \nu \models a(\g{x})$), and $w, \nu \not\models \exists \g{x}(\varphi)$ for any $\g{x}$ and $\varphi$ (i.e., an existential quantification is false, and conversely a universal quantification is true, if the quantified variable ranges over the empty set).

\fi

To improve readability, we will drop $\nu$ from the notation whenever there is no risk of ambiguity---i.e., we will write $w \models \varphi$ to indicate that string $w$ satisfies formula $\varphi$.

An MFO \emph{sentence} is a closed MFO formula. 
Given a sentence $\varphi$, the language $L(\varphi)$ is defined as
\ifemptystrok
\[
L(\varphi) = \{ w \in \Sigma^* \mid w \models \varphi\}.
\]
\fi

We say that a language $L$ is \textit{expressible} in MFO (or \textit{definable} in MFO or \textit{MFO-definable} for short) iff there exists a MFO sentence $\varphi$ such that $L=L(\varphi)$.

\subsection{Examples}
\label{subsec:examples}

The following MFO formula $\varphi_{L_1}$ defines the language $L_1$ made of all strings that start with symbol $a$:
$$
\varphi_{L_1}: \exists \g{x} (\g{x} = 0 \land a(\g{x}))
$$

The following formula $\varphi_{L_2}$ defines the language $L_2$ made of all strings in which every symbol $a$ is necessarily immediately followed by a $b$ (notice that these strings cannot end with a symbol $a$).
$$
\varphi_{L_2}: \forall \g{x} (a(\g{x}) \Rightarrow \exists \g{y} (\sux(\g{x},\g{y}) \land b(\g{y})))
$$


The following formula $\varphi_{L_3}$ defines the language $L_3$ made of all
\ifemptystrok
(nonempty)
\fi
strings in which the last symbol is an $a$.
$$
\varphi_{L_3}: \exists \g{x} (\last(\g{x}) \land a(\g{x}))
$$

The following formula $\varphi_{L_4}$ defines the language $L_4$ made of all strings (containing at least 3 symbols)
in which the symbol three positions from the right is an $a$.
$$
\varphi_{L_4}: \exists \g{x} (a(\g{x}) \land \exists \g{y} (\g{y} = \g{x} + 2 \land \last(\g{y})))
$$
Alternatively, language $L_4$ is also defined by the following formula $\varphi'_{L_4}$:
$$
\varphi'_{L_4}: \exists \g{x} (\last(\g{x}) \land \exists \g{y} (\g{y} = \g{x} - 2 \land a(\g{y})))
$$

\ifemptystrok
The following formula $\varphi_{L_\epsilon}$ (or, alternatively, $\varphi'_{L_\epsilon}$) defines the language $L_\epsilon$ made of only the empty string (assuming that the input alphabet includes at least symbol $a$):
$$
\varphi_{L_\epsilon}: \neg \exists \g{x} (a(\g{x}) \lor \neg a(\g{x}))
$$
$$
\varphi'_{L_\epsilon}: \forall \g{x} (a(\g{x}) \land \neg a(\g{x}))
$$
Formula $\varphi_{L_\epsilon}$ (resp., formula $\varphi'_{L_\epsilon}$) states that, given a word $w$ of language $L_\epsilon$, there does not exist a position such that \emph{true} holds (resp., in all its positions \emph{false} holds).
This can only occur if the set of positions is empty: by definition an existential quantification is false, and a universal quantification is true, if it ranges over an empty set, that is, if $w$ is the empty string.
\fi

Finally, the following formula $\varphi_{L_\emptyset}$ defines the empty language (assuming that the input alphabet $\Sigma$ includes at least symbol $a$):
$$
\varphi_{L_\emptyset}: \exists \g{x} (a(\g{x}) \land \neg a(\g{x}))
$$
Formula $\varphi_{L_\emptyset}$ is contradictory, as it states that a position exists in which symbol $a$ both appears and does not appear.
No string $w$
\ifemptystrok
(not even the empty one)
\fi
is such that $w \models \varphi_{L_\emptyset}$ holds, hence the formula defines the empty language.

Every singleton language---i.e., every language consisting of one finite-length string---is trivially expressible in MFO. Consider, for instance, 
language $L_{abc}= \{ abc  \}$ that includes only string $abc$.
It is easily defined by the following MFO formula:
$$
\varphi_{L_{abc}}: \exists \g{x} \exists \g{y} \exists \g{z} ( \g{x}=0 \land \g{y}=succ(\g{x}) 
 \land \g{z}=succ(\g{y}) \land last(\g{z}) \land 
 a(\g{x}) \land b(\g{y}) \land c(\g{z}) )
$$

\subsection{Expressiveness of MFO}
\label{subsec:MFOexpressiveness}

The following statements trivially hold.

\begin{proposition}
\label{prop:MFOoperations}
Let $L$, $L_1$, and $L_2$ be any languages defined by MFO formulae $\varphi$,  $\varphi_1$ and $\varphi_2$, respectively:
\begin{itemize}

\item
Language $L_1 \cap L_2$ is defined by formula $\varphi_1 \land \varphi_2$---i.e., $L(\varphi_1 \land \varphi_2) = L_1 \cap L_2$.

\item
Language $L_1 \cup L_2$ is defined by formula $\varphi_1 \lor \varphi_2$---i.e., $L(\varphi_1 \lor \varphi_2) = L_1 \cup L_2$.

\item
Language $\bar{L}$ (the complement of $L$) is defined by formula $\neg \varphi$---i.e., $L(\neg \varphi) = \bar{L}$.

\end{itemize}
\end{proposition}

The next theorem follows from Proposition \ref{prop:MFOoperations}.

\begin{theorem}
\label{th:MFOclosure}
The family of MFO-definable languages is closed under union, intersection, and complementation.
\end{theorem}

To further investigate the expressive power of MFO,\footnote{In the present section we follow the line of discussion adopted in the (unedited, to the best of our knowledge) lecture notes \textit{Automata theory - An algorithmic approach} by Javier Esparza, February 13, 2019.} we consider the MFO-definable languages over a one-letter alphabet $\Sigma=\{a\}$. In this simple case the MFO predicate $a(\g{x})$ is always true at any position $x$ in any interpretation, therefore it is redundant and every formula is equivalent to a formula that does not include any occurrence of predicate $a(\cdot)$---e.g., $\exists \g{x}\,(a(\g{x}) \land \g{y}<\g{x})$ is equivalent to $\exists x\,(\g{y}<\g{x})$. 

We next show that, for the simple family of the languages over a one-letter alphabet, every language is MFO-definable if, and only if, it is finite\footnote{Recall that a finite language is one whose cardinality is finite.} or co-finite (where a co-finite language is one whose complement is finite). As a consequence, for instance, the simple \textit{regular} language $L_{even}=\{\, a^{2n} \mid n \geq 0 \,\}$ is \textit{not} MFO-definable, which proves that the MFO logic is strictly less expressive than finite state automata and regular grammars and expressions. 

Our proof that every language over a one-letter alphabet is expressible in MFO if, and only if, it is finite or co-finite is organized as follows. First, we observe that if a language is finite or co-finite, then it is expressible in MFO, as a consequence of the fact that---as exemplified by language $L_{abc}$ in Section \ref{subsec:examples}---singleton languages are expressible in MFO and that the family of MFO-expressible languages is closed under union and complementation. Next, we prove that if a language over a one-letter alphabet is MFO-definable, then it is finite or co-finite. This is in turn proved in three steps: 
\begin{enumerate}
    \item we introduce a new logic called QF-MFO, a quantifier-free fragment of MFO;
    \item we show that every language over a one-letter alphabet $\Sigma = \{ a \}$ is QF-MFO-definable if, and only if, it is finite or co-finite;
    \item we show that the two logics, MFO and QF-MFO, are equally expressive, as every MFO formula $\varphi$ has an equivalent QF-MFO formula.
\end{enumerate}

To define the QF-MFO logic, we first introduce a few additional abbreviations (where $k$ is a constant in $\mathbb{N}$):
%
$$
\begin{aligned}
\g{x} < \g{y}+k           \; \deffmla & \; \ \exists \g{z}\,(\g{z}=\g{y}+k \land \g{x}<\g{z}) \\
\g{x} > \g{y}+k           \; \deffmla & \; \ \exists \g{z}\,(\g{z}=\g{y}+k \land \g{z}<\g{x}) \\
\g{x} < k \; \deffmla & \; \ \exists \g{z}\,(\g{z}=0 \land \g{x}<\g{z} + k) \\
\g{x} > k \; \deffmla & \; \  \exists \g{z}\,(\g{z}=0 \land \g{x} > \g{z} + k) \\
k < last           \; \deffmla & \; \ \forall \g{x}\,(last(\g{x}) \Rightarrow \g{x}>k) \\
k > last           \; \deffmla & \; \ \forall \g{x}\,(last(\g{x}) \Rightarrow \g{x}<k) \\
\end{aligned}
$$
%
\begin{definition}[QF-MFO]
The formulae of QF-MFO are defined by the following syntax:
\[
\varphi :=
           \g{x} < k \; \mid \; \g{x} > k \; \mid \; \g{x} < \g{y} + k  \; \mid \; 
           \g{x} > \g{y} + k  \; \mid \; k < last \; \mid \; 
           k > last \; \mid \; \varphi \land \varphi  \; \mid \; 
           \varphi \lor \varphi
\]
where $\g{x}, \g{y} \in \mathcal{V}$ and $k \in \mathbb{N}$.
\end{definition}

In the remainder, with some (innocuous) overloading, a constant $k$ will denote both the numerical value $k \in \mathbb{N}$ and the string $a^k$.

\begin{proposition}
\label{QF-FIN-COFIN}
Every language $L$ over a one-letter alphabet is QF-MFO-definable if, and only if, it is finite or co-finite. 
\end{proposition}

\begin{proof} \textit{Only if part}: Every QF-MFO sentence defines a finite or a co-finite language.
\par
Let $\varphi$ be a sentence of QF-MFO. Since QF-MFO is quantifier-free, the sentence $\varphi$ is an and-or combination of formulae of type $k<last$ and $k>last$. Then, the following cases arise. 
\begin{itemize}
    \item $L(k<last)=\{k+1, k+2,\;\ldots\;\}$ is a co-finite language (remember that numbers identify words, and vice versa, so that $\{k+1, k+2,\;\ldots\;\}$ is the same as $\{a^{k+1}, a^{k+2},\;\ldots\;\}$).
    \item $L(k>last)=\{0, 1,\;\ldots\;k\}$ is a finite language.
    \item $L(\varphi_1 \lor \varphi_2)=L(\varphi_1) \cup L(\varphi_2)$. If $L_1=L(\varphi_1)$ and $L_2=L(\varphi_2)$ are both finite, then $L(\varphi_1 \lor \varphi_2)$ is also finite; if $L_1$ and $L_2$ are both co-finite, then the language $L(\varphi_1 \lor \varphi_2)=L_1 \cup L_2 = \overline{ \overline{L_1 \cup L_2} } = \overline{ \overline{L_1} \cap \overline{L_2} }$ is the complement of the intersection of two finite languages, hence it is co-finite; if one of the two languages $L_1$ and $L_2$ is finite and the other is co-finite, then $L(\varphi_1 \lor \varphi_2)$
    is the complement of the intersection of a finite and a co-finite language, therefore it is co-finite. 
    \item $L(\varphi_1 \land \varphi_2)=L(\varphi_1) \cap L(\varphi_2)$. If $L_1=L(\varphi_1)$ and $L_2=L(\varphi_2)$ are both finite, then their intersection is finite; if $L_1$ and $L_2$ are both co-finite, then $L(\varphi_1 \land \varphi_2)=L_1 \cap L_2 = \overline{ \overline{L_1 \cap L_2} } = \overline{ \overline{L_1} \cup \overline{L_2} }$ is the complement of the union of two finite languages, hence it is co-finite; if one of the two languages $L_1$ and $L_2$ is finite and the other is co-finite, then $L(\varphi_1 \land \varphi_2)$ is the complement of the union of a finite language and a co-finite language, hence it is finite.  
\end{itemize}

\par
\textit{If part}: Every finite or co-finite language is definable by a QF-MFO sentence.
\par
If $L$ is finite then $L=\{k_1,\;\ldots\;k_n\}$ and 
$$
\varphi_L = \varphi_{\{k_1\}} \lor \; \cdots \; \lor \; \varphi_{\{k_n\}} =
$$
$$
= (k_1-1<last \land last<k_1+1)\;\lor\;\cdots\;\lor\;(k_n-1<last \land last<k_n+1)
$$
If $L$ is co-finite, then its complement $\overline{L}$ is finite, therefore it is defined by some QF-MFO formula. 
Then, $L$ is QF-MFO-definable if, for every QF-MFO sentence $\varphi$, there exists a QF-MFO sentence, call it $\overline{\varphi}$, that defines the language $\overline{L}$, the complement of $L$. Such a sentence $\overline{\varphi}$ is equal to $neg(\varphi)$, where function $neg(\cdot)$ is defined inductively by the following clauses.
\begin{itemize}
    \item $neg(k<last) = last<k \lor \underbrace{(k-1<last \land last<k+1)}_{last = k}$
    \item $neg(k>last) = k<last \lor \underbrace{(k-1<last \land last<k+1)}_{k = last}$
    \item $neg(\varphi_1 \lor \varphi_2)=neg(\varphi_1) \land neg(\varphi_2)$
    \item $neg(\varphi_1 \land \varphi_2)=neg(\varphi_1) \lor neg(\varphi_2)$
\end{itemize}
\qed
\end{proof}

\begin{proposition}
\label{MFO-EQ-QF}
Every MFO formula $\varphi$ over a one-letter alphabet is equivalent to some QF-MFO formula $f$---i.e., $\varphi \equiv f$.
\end{proposition}

\begin{proof}
The proof is by induction on the structure of $\varphi$.

If $\varphi=\g{x}<\g{y}$, then $\varphi\equiv \g{x}<\g{y}+0$.

If $\varphi = \neg \psi$, then the inductive hypothesis can be applied and then the negation can be removed using the De Morgan's laws and equivalences such as, e.g., $\neg(\g{x}<\g{y}+k)\equiv \g{x} \geq \g{y}+k$ (where $\g{x} \geq \g{y}+k$ is a natural abbreviation for $\g{x} > \g{y}+k-1$).

If $\varphi = \varphi_1 \lor \varphi_2$, then the induction hypothesis is directly applied. 

If $\varphi = \exists \g{x} \, \psi$ then, by the induction hypothesis, $\psi \equiv f$ for some QF-MFO formula $f$, and $f$ can be assumed to be in disjunctive normal form---i.e., $f=D_1 \lor \cdots \lor D_n$, and $\varphi \equiv \exists \g{x} D_1 \lor \cdots \lor \exists \g{x} D_n$; then, we define a set of QF-MFO formulae $f_i$ such that, for each $1 \leq i \leq n$, $f_i \equiv  \exists\,\g{x} D_i$ holds.
Notice that, since $f$ is a QF-MFO formula, each $f_i$ is such that it does not include any quantification of variable $\g{x}$ nor, if $\varphi$ is a sentence, any occurrence of the same variable.  

Each $f_i$ is built as follows. Formula $f_i$ is a conjunction of formulae that contains all the conjuncts of $D_i$ that do not include any occurrence of variable $\g{x}$, plus the conjuncts defined next. Consider every pair of conjuncts of $D_i$, one conjunct being of type $t_1<\g{x}$, where $t_1=h$ or $t_1=\g{y}+h$
and the constraint is an \textit{upper bound} (i.e., $h$ is maximal, that is, it is the greatest value that appears in a constraint of the type $\g{y}+h <\g{x}$),
and the other conjunct being of type $\g{x}<t_2$, where $t_2=h$ or $t_2=\g{y}+h$ and the constraint is a \textit{lower bound} (i.e., $h$ is minimal); for every such pair we add to $f_i$ a conjunct equivalent to $t_1+1<t_2$; for instance, if the two above-described conjuncts are $\g{z}-4<\g{x}$ and $\g{x}<\g{y}+3$, then the added conjunct is $\g{z}
<\g{y}+6 \equiv \g{z}-3<\g{y}+3$. Notice that, if only the conjunct of type $t_1<\g{x}$ is present and the conjunct of type $\g{x}<t_2$ is missing, then the (trivially true) conjunct $\g{x}<last+1$ must be considered---in place of the latter---as the other element of the pair; similarly, if only the conjunct of type $\g{x}<t_2$ is present and the conjunct of type $t_1<\g{x}$ is missing, then the (trivially true) conjunct $-1<\g{x}$ must be considered in place of the latter.       

Then, $f_i \equiv \exists \g{x} D_i$; notice that $f_i$ does not include any occurrence of variable $\g{x}$ nor any quantification of that variable.  
\qed
\end{proof}

\begin{example}
In the MFO formula
        $$
        \exists \g{x} (\g{x}<\g{y}+3 \land \g{z}<\g{x}+4 \land \g{z}<\g{y}+2 \land \g{y}<\g{x}+1 )
        $$
we identify the pair of constraints $\g{z}-4<\g{x}$ and $\g{x}<\g{y}+3$, from which we get the additional conjunct $\g{z}-3<\g{y}+3 \equiv \g{z}<\g{y}+6$; we also identify the pair of constraints $\g{y}-1<\g{x}$ and $\g{x}<\g{y}+3$, from which we get the additional conjunct $\g{y}<\g{y}+3$. Therefore, the MFO formula $\exists \g{x} (\g{x}<\g{y}+3 \land \g{z}<\g{x}+4 \land \g{z}<\g{y}+2 \land \g{y}<\g{x}+1 )$ is equivalent to the QF-MFO formula 
        $$
        \g{z}<\g{y}+6 \land \g{y}<\g{y}+3 \land \g{z}<\g{y}+2
        $$
\qed
\end{example}
\begin{example} We provide two examples of QF-MFO formulae equivalent to given MFO sentences. 
\par
\begin{itemize} 
    \item The MFO formula $\exists \g{x}\,\exists \g{y}\,\exists \g{z}\,(\,\g{x}<\g{y} \land \g{y}<\g{z}\,)$ defines the language $\{\, a^k \mid k \geq 3 \,\}$. By repeated application of the inductive step, moving inside-out, we obtain $f_1 \equiv  \underbrace{\exists \g{z}\,(\,\g{x}<\g{y} \land \g{y}<\g{z}\,)}_{\exists \g{z}\;D_1}$ and the pair of constraints on the quantified variable $\g{z}$, $\g{y}<\g{z}$ and $\g{z}<last +1$, from which we derive constraint $\g{y}<last$, so that $f_1 \equiv \g{x}<\g{y} \land \g{y}<last$ holds; in the next inductive step we have $f_1 \equiv \underbrace{\exists\, \g{y}(\g{x}<\g{y} \land \g{y}<last)}_{\exists \g{y}\; D_1}$ and the pair of constraints on the quantified variable $\g{y}$, $\g{x}<\g{y}$ and $\g{y}<last$, from which we derive $\g{x}+1<last$ and $f_1 \equiv \g{x}+1<last$; in the final inductive step we have $f_1 \equiv \underbrace{\exists\, \g{x}\,(\g{x}+1 <last)}_{\exists \g{x}\; D_1}$ and the pair of constraints on the quantified variable $\g{x}$, $-1<\g{x}$ and $\g{x}<last-1$, from which we obtain $0<last-1$ and $f_1 \equiv last>1$, hence $last>1$ is the QF-MFO formula equivalent to the original MFO formula $\exists \g{x}\,\exists y\,\exists z\,(\,\g{x}<\g{y} \land \g{y}<\g{z}\,)$.
    
    \item The MFO formula $\exists \g{x}\,(\,\neg \exists \g{y}\,(\g{x}<\g{y}) \land \g{x}<4\,)$ defines the language $\{\, a^k \mid k \leq 4 \,\}$.
    Again moving inside-out, we have $f_1 \equiv \underbrace{\exists \g{y}\,(\,\g{x}<\g{y} \,)}_{\exists \g{y}\;D_1}$ and the pair of constraints on the quantified variable $\g{y}$, $\g{x}<\g{y}$ and $\g{y}<last+1$, from which we derive $\g{x}+1<last+1 \equiv \g{x}<last$ and $f_1 \equiv \g{x}<last$; at the next inductive step we apply negation and obtain $f_1 \equiv  \underbrace{\exists \g{x}\,(\,\g{x} \geq last \land \g{x}<4 \,)}_{\exists \g{x}\;D_1}$ and the pair of constraints on the quantified variable $\g{x}$, $last-1<\g{x}$ and $\g{x}<4$, from which we obtain $f_1 \equiv last<4$. Hence, $last<4$ is the QF-MFO formula equivalent to the original MFO formula $\exists \g{x}\,(\,\neg \exists \g{y}\,(\g{x}<\g{y}) \land \g{x}<4\,)$.  
\end{itemize}
\qed
\end{example}

The following theorem easily follows from Proposition \ref{QF-FIN-COFIN} and Proposition \ref{MFO-EQ-QF}.

\begin{theorem}
\label{prop:MFOfinOrCofin}
A language over a one-letter alphabet is expressible in MFO if, and only if, it is finite or co-finite.
\end{theorem}

%
%
%
%
%
Every MFO formula defines a regular language.
In fact, MFO is a restriction of the Monadic Second-Order (MSO) logic introduced in Section \ref{sec:MSO} and, as shown there, for every MSO sentence $\varphi$ there is a Finite-State Automaton (FSA) that accepts exactly the language defined by $\varphi$---hence, \emph{a fortiori} this also holds for every MFO formula.
We have therefore the following result (whose proof will be given in Section \ref{subsec:MSOexpressiveness}).
\begin{statement}
\label{prop:MFOtoFSA}
For every MFO sentence $\varphi$ there is a FSA $\mathcal{A}$ such that $L(\varphi) = L(\mathcal{A})$.
\end{statement}
However, the MFO logic is strictly less expressive than Finite State Automata (also abbreviated as FSA), as not all regular languages are expressible in MFO. Indeed,
as a consequence of Theorem \ref{prop:MFOfinOrCofin} the regular language $L_{\text{even}}$ defined above, which includes exactly the strings over alphabet $\Sigma = \{a\}$ having even length and therefore is neither finite nor co-finite, is $not$ expressible in MFO, as stated by the next corollary.
\begin{corollary}
\label{prop:MFOnoLeven}
There is no MFO sentence $\varphi$ that defines language $L_{\text{even}}$ (i.e., such that $L(\varphi) = L_{\text{even}}$).
\end{corollary}

From Statement \ref{prop:MFOtoFSA} and Corollary \ref{prop:MFOnoLeven} the following result is immediate, by observing that it is easy to define a FSA $\mathcal{A}$ such that $L(\mathcal{A}) = L_{\text{even}}$.

\begin{theorem}
\label{th:MFOexpressiveness}
MFO is strictly less expressive than FSA.
\end{theorem}

On the other hand, the set of languages that can be defined through MFO formulae is not closed under the so-called ``Kleene star'' operation, as stated by the following theorem.

\begin{theorem}
\label{th:MFOnonclosure}
The set of languages that can be defined by MFO sentences is not closed under the ${}^*$ operation.
\end{theorem}

\begin{proof}
To prove the claim it is enough to remark that the following MFO formula $\varphi_{L_{aa}}$ defines language $L_{aa} = \{aa\}$ (i.e., the language containing only string $aa$):
$$
\varphi_{L_{aa}} : \exists \g{x} \exists \g{y} ( \g{x} = 0 \land \g{y} = \g{x} + 1 \land a(\g{x}) \land a(\g{y}) \land \last(\g{y}) )
$$
and that $L_{\text{even}} = L_{aa}^*$.
\qed
\end{proof}

From 
Theorem \ref{th:MFOclosure} and Theorem \ref{th:MFOnonclosure},
it can be shown \cite{McNauPap} that MFO can express only the so-called ``star-free'' languages---that is, those that can be obtained through union, intersection, complementation and concatenation of finite languages.

\ifemptystrok
For example, language $L_3$ of Section \ref{subsec:examples} can be obtained from finite languages $L_3' = \emptyset$ and $L_3'' = \{a\}$---containing, respectively, no string (hence, whose cardinality is $0$) and only string $a$ (hence, whose cardinality is $1$)---in the following way:
$$
L_3 = (\neg L_3') \cdot L_3''
$$
As a further example, the language $L_{\exists a}$, which is made of all strings that contain at least an $a$, can be defined in the following way:
$$
L_{\exists a} = (\neg L_3') \cdot L_3'' \cdot (\neg L_3')
$$
and the language $L_{\exists ! a}$ made of all strings that contain \emph{exactly} one $a$ can be defined as follows:
$$
L_{\exists ! a} = (\neg L_{\exists a}) \cdot L_3'' \cdot (\neg L_{\exists a})
$$
\fi

%% file: mso.tex
\section{Monadic Second-order Logic of Order on Strings}
\label{sec:MSO}


Formulae of the \emph{monadic second-order logic of order} (MSO), as defined by B\"uchi and others \cite{bib:Thomas1990a},
are built out of the elements of the MFO logic defined in Section \ref{sec:MFO} plus, in addition:

\begin{itemize} 

\item
Second-order variables, denoted as uppercase boldface letters, $\bm X$, $\bm Y$, \ldots, which are interpreted over \emph{sets} of natural numbers.

%

\end{itemize}

More precisely, let $\Sigma$ be an input alphabet, $\mathcal{V}_1$ be a set of first-order variables, and  $\mathcal{V}_2$ be a set of second-order (or set) variables.
Well-formed formulae of MSO logic are defined according to the following syntax:
\[
\varphi :=
           a(\bm x)
             \; \mid \;
           \bm X(\bm x)
             \; \mid \;
           \g{x} < \g{y} 
             \; \mid \;
           \neg \varphi
             \; \mid \;
           \varphi \lor \varphi 
             \; \mid \;
           \exists \bm x (\varphi)
             \; \mid \;
           \exists \bm X (\varphi)
\]
where $a \in \Sigma$, $\g{x}, \g{y} \in \mathcal{V}_1$, and $\g{X} \in \mathcal{V}_2$.


%

Naturally, all abbreviations introduced in Section \ref{sec:MFO} are still valid.
We also introduce the following additional abbreviations:
$$
\begin{aligned}
\g{x} \in \g{X}           \; \deffmla & \; \ \g{X}(\g{x})     \\
\g{X} \subseteq \g{Y}     \; \deffmla & \; \ \forall \g{x} (\g{x} \in \g{X} \Rightarrow \g{x} \in \g{Y}) \\
\g{X} = \g{Y}             \; \deffmla & \; \ (\g{X} \subseteq \g{Y}) \land (\g{Y} \subseteq \g{X})  \\
\g{X} \neq \g{Y}          \; \deffmla & \; \ \neg (\g{X} = \g{Y})
\end{aligned}
$$
where $\g{x}, \g{y}, \g{X}$ are as before, and $\g{Y} \in \mathcal{V}_2$.

An MSO formula is interpreted over a string
\ifemptystrok
$w \in \Sigma^*$,
\fi
with respect to assignments $\nu_1: \mathcal{V}_1 \to \{0, \ldots, |w|-1\}$ and $\nu_2: \mathcal{V}_2 \to \powerset(\{0, \ldots, |w|-1\})$.
Notice that, like assignment $\nu$ for MFO formulae, $\nu_1$ maps each first-order variable of $\mathcal{V}_1$ to a position in string $w$.
Assignment $\nu_2$, instead, maps each second-order variable of $\mathcal{V}_2$ to a \emph{set} of positions in string $w$.

Then, the satisfaction relation $\models$ for MSO formulae is defined in the following way:

\ifemptystrok

\begin{itemize}
    \item
    $w, \nu_1, \nu_2 \models a(\g{x})$ if, and only if, $\nu_1(\g{x})$ is defined and $w[\nu_1(\g{x})]=a$ 
    \item
    $w, \nu_1, \nu_2 \models \g{X}(\g{x})$ if, and only if, $\nu_1(\g{x}) \in \nu_2(\g{X})$ holds 
    \item
    $w, \nu_1, \nu_2 \models \g{x} < \g{y}$ if, and only if, $\nu_1(\g{x}) < \nu_1(\g{y})$ holds
    \item
    $w, \nu_1, \nu_2 \models \neg\varphi$ if, and only if, $w, \nu_1, \nu_2 \not\models \varphi$ holds
    \item
    $w, \nu_1, \nu_2 \models \varphi_1 \lor \varphi_2$ if, and only if, at least one of $w, \nu_1, \nu_2 \models \varphi_1$ and  $w, \nu_1, \nu_2 \models \varphi_2$ holds
    \item
    $w, \nu_1, \nu_2 \models \exists \g{x}(\varphi)$ if, and only if, $|w|>0$ and some $i \in \{0,\,\ldots,\,|w|-1 \}$ satisfies $w, \nu_1[i / \g{x}], \nu_2 \models \varphi$
    \item
    $w, \nu_1, \nu_2 \models \exists \g{X}(\varphi)$ if, and only if, $|w|>0$ and some $S\subseteq \{\,0\,\ldots\,|w|-1 \,\}$ satisfies  $w, \nu_1, \nu_2[S/\g{X}] \models \varphi$
\end{itemize}
where $w[k]$ and $\nu[i / \g{x}]$ are as in Section \ref{sec:MFO}, and $\nu_2[S/\g{X}]$ assigns $S$ to $\g{X}$ and otherwise coincides with $\nu_2$.

\fi

To improve readability, we will drop $\nu_1$, $\nu_2$ from the notation whenever there is no risk of ambiguity, and write $w \models \varphi$ to indicate that string $w$ satisfies MSO formula $\varphi$.

The definitions of MSO \emph{sentence} and of language $L(\varphi)$ defined by sentence $\varphi$ are as for the MFO logic.


\begin{example}
\label{ex:MSOLeven}
The following MSO formula $\varphi_{\text{even}}$ defines language $L_{\text{even}}$ introduced in Section \ref{subsec:MFOexpressiveness}.
$$
\varphi_{\text{even}} :
\exists \g{P}
  \left(
    \forall \g{x}
      \left(
        \begin{array}{l}
                ( \g{x} = 0 \Rightarrow \neg \g{P}(x) )
                \\ \land \\
                \forall \g{y} ( \g{y} = \g{x} + 1 \Rightarrow (\neg \g{P}(\g{x}) \Leftrightarrow \g{P}(\g{y})) )
                \\ \land \\
                a(\g{x})
                \\ \land \\
                ( \last(\g{x}) \Rightarrow \g{P}(\g{x}) )
        \end{array}
      \right)
  \right)
$$
Formula $\varphi_{\text{even}}$ introduces a second-order variable $\g{P}$ that identifies exactly all even positions in string $w$.
More precisely, the first position of $w$ (which is conventionally $0$), is not even, and indeed the first conjunct in formula $\varphi_{\text{even}}$ states that $\g{P}(x)$ does not hold when $\g{x}$ is $0$.
In addition, the second conjunct in $\varphi_{\text{even}}$ states that the next position after $\g{x}$ (i.e., position $\g{y}$ such that $\g{y}=\g{x}+1$ holds), if it exists, is even (i.e., $\g{P}(\g{y})$ holds there) if, and only if, position $\g{x}$ is odd;
hence, since the first position is odd, the second position is even, the third is odd, the fourth is even, and so on.
The third conjunct states that, in every position $\g{x}$, $a(\g{x})$ holds (i.e., $a$ appears in every position).
Finally the last conjunct states that the last position in the string must be even.
\qed
\end{example}

\subsection{Expressiveness of MSO}
\label{subsec:MSOexpressiveness}

Since every MFO formula is also an MSO formula, from the fact that MSO formula $\varphi_{\text{even}}$ introduced in Example \ref{ex:MSOLeven} defines language $L_{\text{even}}$, which, by Proposition \ref{prop:MFOnoLeven}, cannot be defined by an MFO formula, we have the following straightforward result.

\begin{theorem}
\label{th:MSOgtMFO}
The MSO logic is strictly more expressive than the MFO logic.
\end{theorem}

Indeed, the original seminal result by B\"uchi and others is that, unlike the MFO logic, MSO indeed has the same expressive power as FSAs, as captured by the following theorem.

\begin{theorem}
\label{th:RLequivMSO}
A language $L$ is regular if, and only if, there exists a sentence $\varphi$ in the MSO logic such that $L = L(\varphi)$.
\end{theorem}

Before proving Theorem \ref{th:RLequivMSO}, we remark that, since for every FSA there is an equivalent MSO formula---and vice versa---MSO enjoys all closure properties of FSAs, as captured by the following corollary.

\begin{corollary}
\label{cor:MSOclosure}
The set of languages that can be defined by MSO formulae is closed under union, intersection, complementation, and Kleene star.
\end{corollary}

The proof of Theorem \ref{th:RLequivMSO} is constructive, i.e., it provides an algorithmic procedure that, for a given FSA $\mathcal{A}$, builds an equivalent MSO sentence $\varphi_{\mathcal{A}}$, and vice versa.
Next, we offer an intuitive explanation of the construction, referring the reader to, e.g., \cite{bib:Thomas1990a} for a complete and
detailed proof.  

\subsection*{From FSA to MSO logic}

The key idea of the construction consists in using, for each  state $q$ of FSA $\mathcal{A}$, a second-order variable $\bm{X}_q$, whose value is the set of positions of all the characters that $\mathcal{A}$ may read in a transition starting from state $q$.

Without loss of generality, we assume that $\mathcal{A}$'s set of states $Q$ is $\{0, 1, \ldots, m\}$, for some $m$, where $0$ denotes the initial state.
Then, we encode the definition of the FSA $\mathcal{A}$ recognizing $L$ (i.e, such that $L = L(\mathcal{A})$) as the conjunction of several clauses, each one capturing a part of the definition of $\mathcal{A}$:

\begin{itemize}
\item
We introduce a formula capturing the transition relation $\delta$ of $\mathcal{A}$, which includes a disjunct for each transition $\delta(q_i,a) = q_j$ of the automaton:\\
$\forall \g{x}, \g{y}\left(\g{y} = \g{x} +1 \Rightarrow \bigvee_{\delta(q_i,a) = q_j } \left(\g{x} \in \g{X}_i \land a(\g{x}) \land  \g{y} \in \g{X}_{j}\right)\right)$.


\item
The fact that the machine starts in state $0$ is captured by formula\\
$\forall \bm x (\g{x} = 0 \Rightarrow \bm x \in \bm X_{0})$.

\item
Since the automaton cannot be in two different states $i, j$ at the same time, for each pair of distinct second-order variables $\bm X_i$ and $\bm X_j$ we introduce formula\\
$\neg\exists \bm y (\bm y \in \bm X_i \land \bm y \in \bm X_j)$. 

\item
Acceptance by the automaton---i.e. $\delta(q_i,a) \in F$---is formalized by formula\\
$\forall \g{x}\left( \last(\g{x}) \Rightarrow \bigvee_{\delta(q_i,a) \in F} \left(\g{x} \in \bm X_i \land a(\g{x})\right) \right)$.

\end{itemize}
%
Finally, MSO formula $\varphi_{\mathcal{A}}$ corresponding to automaton $\mathcal{A}$ is the following sentence
$$\varphi_{\mathcal{A}} : \exists \bm X_0, \bm X_1, \ldots \bm X_m (\varphi)$$
where $\varphi$ is the conjunction of all the above clauses.

It is not difficult to show that the set of strings satisfying formula $\varphi_{\mathcal{A}}$ is exactly $L$.

\begin{example}
\label{ex:FSAtoMSO}

Consider the FSA $\mathcal{A}_\text{ex}$ shown in Figure \ref{fig:exampleFSA}.
\begin{figure*}
  \centering
  \includegraphics[scale=0.7]{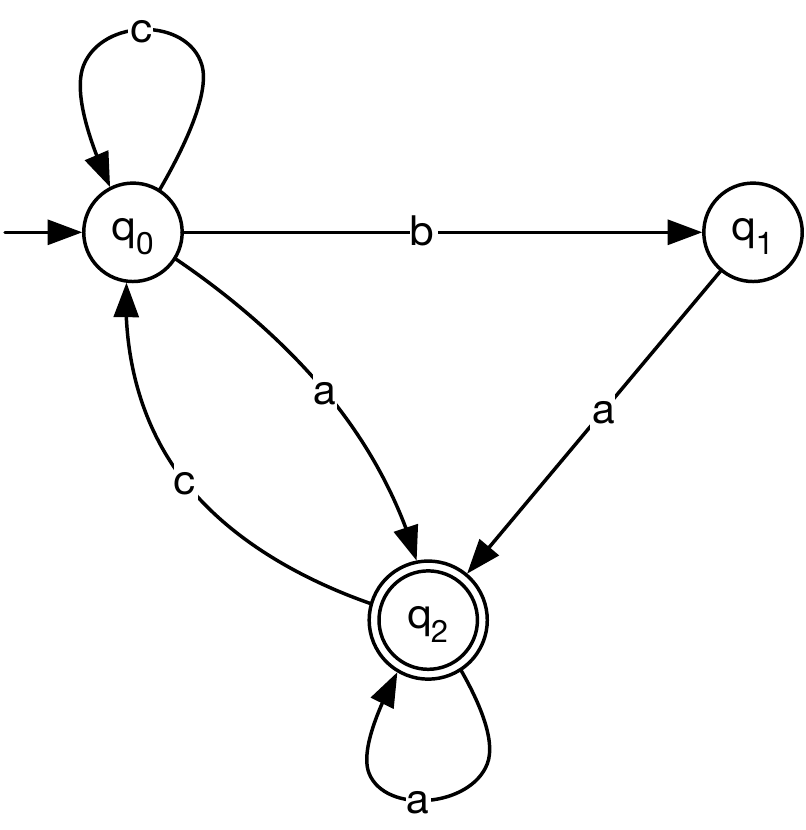}
  \caption{Finite-State Automaton $\mathcal{A}_\text{ex}$.}
  \label{fig:exampleFSA}
\end{figure*}
The corresponding MSO formula $\varphi_{\mathcal{A}_\text{ex}}$ built according to the rules described above is the following:
%
$$
\varphi_{\mathcal{A}_\text{ex}} :
\exists \g{X}_0, \g{X}_1, \g{X}_2
  \left(
  \begin{array}{l}
     \forall \g{x}, \g{y}\left(\g{y} = \g{x} +1 \Rightarrow
                                    \left( 
                                      \begin{array}{l}
                                        \g{x} \in \g{X}_0 \land c(\g{x}) \land  \g{y} \in \g{X}_{0} \quad \lor \\
                                        \g{x} \in \g{X}_0 \land b(\g{x}) \land  \g{y} \in \g{X}_{1} \quad \lor \\
                                        \g{x} \in \g{X}_0 \land a(\g{x}) \land  \g{y} \in \g{X}_{2} \quad \lor \\
                                        \g{x} \in \g{X}_1 \land a(\g{x}) \land  \g{y} \in \g{X}_{2} \quad \lor \\
                                        \g{x} \in \g{X}_2 \land c(\g{x}) \land  \g{y} \in \g{X}_{0} \quad \lor \\
                                        \g{x} \in \g{X}_2 \land a(\g{x}) \land  \g{y} \in \g{X}_{2}
                                      \end{array}
                                    \right) \right)
     \\ \land \\
     \forall \bm x (\g{x} = 0 \Rightarrow \bm x \in \bm X_{0}) \
     \\ \land \\
     \neg\exists \bm y (\bm y \in \bm X_0 \land \bm y \in \bm X_1) \
     \land \\
     \neg\exists \bm y (\bm y \in \bm X_0 \land \bm y \in \bm X_2) \
     \land \\
     \neg\exists \bm y (\bm y \in \bm X_1 \land \bm y \in \bm X_2) \
     \\ \land \\
     \forall \g{x}\left( \last(\g{x}) \Rightarrow
                              \left(
                              \begin{array}{l}
                                (\g{X}_0(\g{x}) \land a(\g{x}))
                                \\ \lor \\
                                (\g{X}_1(\g{x}) \land a(\g{x}))
                                \\ \lor \\
                                (\g{X}_2(\g{x}) \land a(\g{x}))
                              \end{array}
                              \right)
                  \right)
     
  \end{array}
  \right)
$$
where the first clause captures the transition relation of the automaton; the second clause formalizes its initial state; the next three conjuncts state the mutual exclusion of states; and the last clause captures the acceptance condition.
\qed
\end{example}

\ifemptystrok
\begin{remark}
The formalization of FSA $\mathcal{A}$ through MSO formula $\varphi_{\mathcal{A}}$, and in particular the clause capturing the acceptance condition, assumes that accepted strings contain at least one symbol (i.e., they are not empty).
To formalize in MSO an FSA $\mathcal{A_\epsilon}$ that also accepts the empty string, it is enough to include in the corresponding formula $\varphi_{\mathcal{A_\epsilon}}$ also a disjunct covering the case of the empty string, in the following way:
$$
\varphi_{\mathcal{A_\epsilon}} : \exists \bm X_0, \bm X_1, \ldots \bm X_m (\varphi) \vee \varphi_{L_\epsilon}
$$
where $\varphi_{L_\epsilon}$ is the formula introduced in Section \ref{subsec:examples} to capture the empty string.
\end{remark}
\fi

\subsection*{From MSO logic to FSA}

The construction in the opposite direction has been
proposed in various versions in the literature. Here we summarize its main
steps along the lines of \cite{bib:Thomas1990a}. 
First, the MSO sentence is translated into a
standard form that uses only second-order variables (no first-order variables are allowed), the $\subseteq$ predicate, and
variables $\bm W_a$, for each $a \in \Sigma$, denoting the set of all the positions of the word
containing the character $a$.
Moreover, we use $\Sux$, which has the same meaning as $\sux$,
has second-order variable arguments that are singletons. 
This simpler (yet equivalent) logic is defined by the following syntax:
        \[
\varphi := \bm X \subseteq \bm W_a \mid \bm X \subseteq \bm Y \mid \Sux(\bm X, \bm Y) 
\mid \neg \varphi \mid \varphi \lor \varphi 
\mid \exists \bm X (\varphi).
\]
As before, we also use the standard abbreviations for, e.g., $\land$, $\forall$, $=$.
To translate first-order variables to second-order variables we need to
  state that a (second-order) variable is a singleton. Hence we introduce the
  abbreviation:
$$
\begin{aligned}
\Sing(\bm X)  \; \deffmla & \; \ \exists \bm Y ( \bm Y \subseteq \bm X 
\land \bm Y \ne \bm X 
\land \neg\exists \bm Z (\bm Z \subseteq \bm X \land \bm Z \ne \bm Y \land \bm
Z \ne \bm X)
)\\
\end{aligned}
$$
Then, in the transformation below, $\Sux(\bm X, \bm Y)$ is always conjoined with $\Sing(\bm X) \land \Sing(\bm Y)$ and the resulting formula is therefore false whenever $\bm X$ or $\bm Y$ are not singletons.
The following step entails the inductive construction of the equivalent
automaton. This 
is built by
associating a single automaton to each elementary subformula and by composing
them according to the structure of the global formula.
This inductive approach requires to use open formulas, i.e., formulas where free variables occur.
For technical reasons, with such formulas we are going to consider words on the alphabet $\Sigma \times \{0, 1\}^k$, where $k$ is the number of free variables; in the subsequent steps of the transformation from MSO logic to FSA, the alphabet will revert to $\Sigma$.
Hence, if $\bm X_1$, $\bm X_2$,
\ldots $\bm X_k$ are the free variables used in the formula,
a value of 1 in the,
say, $j$-th component means that the considered position belongs to $\bm X_j$ (that is, the second-order variable $\bm X_j$ represents a first-order variable whose value is the considered position); 0 means the opposite. For instance, if $w = (b,1,0)(a,0,0)(a,0,1)$, then $w \models 
\bm X_2 \subseteq \bm W_a$, $w \models \bm X_1 \subseteq \bm W_b$, with $\bm X_1$ and $\bm X_2$ singletons representing (first-order variables and hence) positions in string $w$ respectively equal to $0$ and $2$.

\subsubsection*{Formula transformation}

\begin{enumerate}
\item First order variables are translated in the following way:
$\exists \bm x (\varphi(\bm x))$ becomes\linebreak $\exists \bm X ( \Sing(\bm X) \land \varphi'(\bm X))$,
where $\varphi'$ is the translation of $\varphi$, and $\bm X$ is a fresh new variable not occurring elsewhere.

\item Subformulas having the form $a(\bm x)$, $\sux(\bm x, \bm y)$ are
  translated into $\bm X \subseteq \bm W_a$, $\Sux(\bm X, \bm Y)$, respectively. 
  \item The other parts are unchanged.
\end{enumerate}

\subsubsection*{Inductive construction of the automaton}

We assume for simplicity that $\Sigma = \{a, b\}$, and that $k = 2$, i.e. two
variables are used in the formula. Moreover, in the transition labels of the automata we use the shortcut symbol $\circ$ to mean
all possible values. 

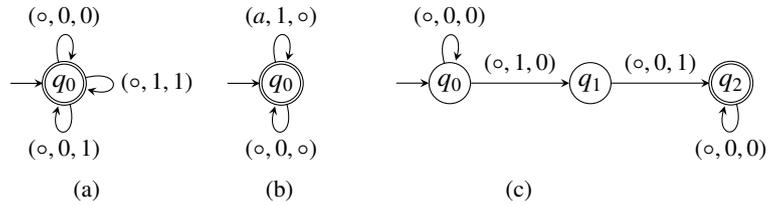
\begin{figure*}
  \centering
\begin{tabular}{m{0.2\textwidth}m{0.2\textwidth}m{0.3\textwidth}}
  \begin{tikzpicture}[every edge/.style={draw,solid}, node distance=3cm, auto, 
                    every state/.style={draw=black!100,scale=0.5}, >=stealth]

\node[initial by arrow, initial text=,state,accepting] (q0) {{\huge $q_0$}};

\path[->]
(q0) edge [loop above] node {$(\circ,0,0)$} (q1)
(q0) edge [loop below] node {$(\circ,0,1)$} (q1)
(q0) edge [loop right] node {$(\circ,1,1)$} (q1);
\end{tikzpicture}
&
\quad
\begin{tikzpicture}[every edge/.style={draw,solid}, node distance=3cm, auto, 
                    every state/.style={draw=black!100,scale=0.5}, >=stealth]

\node[initial by arrow, initial text=,state,accepting] (q0) {{\huge $q_0$}};

\path[->]
(q0) edge [loop below] node {$(\circ,0,\circ)$} (q1)
(q0) edge [loop above] node {$(a,1,\circ)$} (q1);
\end{tikzpicture}
&
\begin{tikzpicture}[every edge/.style={draw,solid}, node distance=3.7cm, auto, 
                    every state/.style={draw=black!100,scale=0.5}, >=stealth]

\node[initial by arrow, initial text=,state] (q0) {{\huge $q_0$}};
\node[state] (q1) [right of=q0] {{\huge $q_1$}};
\node[state] (q2) [right of=q1, accepting] {{\huge $q_2$}};

\path[->]
(q0) edge [loop above]  node {$(\circ,0,0)$} (q0)
(q0) edge [above] node {$(\circ,1,0)$} (q1)
(q1) edge [above] node {$(\circ,0,1)$} (q2)
(q2) edge [loop below]  node {$(\circ,0,0)$} (q2);
\end{tikzpicture}
  \\
  \centering{(a)} 
  &
  \centering{(b)} 
  &
  \centering{(c)} 
\end{tabular}
  \caption{Automata for the construction from MSO logic to FSA.}
  \label{fig:inductive-automata}
\end{figure*}

\begin{itemize}
\item The formula $\bm X_1 \subseteq \bm X_2$ is translated into an automaton
  that checks that there are 1's for the $\bm X_1$ component only in positions where there
  are also 1's for the $\bm X_2$ component (Figure~\ref{fig:inductive-automata}~(a)).

\item The formula $\bm X_1 \subseteq \bm W_a$ is analogous: the automaton checks that
  positions marked by 1 in the $\bm X_1$ component must have symbol $a$ (Figure~\ref{fig:inductive-automata}~(b)).

\item The formula $\Sux(\bm X_1,\bm X_2)$ considers two singletons, and checks
  that the 1 for component $\bm X_1$ is immediately followed by a 1 for
  component $\bm X_2$ (Figure~\ref{fig:inductive-automata}~(c)).

\item Formulas inductively built with $\neg$ and $\lor$ are covered by the closure of regular languages w.r.t.
  complement and union, respectively.
  
\item For a formula of type $\exists \bm X (\varphi)$, we use the closure under
  alphabet projection; for instance, we may start with an automaton with input alphabet $\Sigma
  \times \{0, 1\}^2$, for the formula $\varphi(\bm X_1,\bm X_2)$ and we may need to
  define an automaton for the formula $\exists \bm X_1 (\varphi(\bm X_1, \bm
  X_2))$. But in this case the alphabet is $\Sigma \times \{0, 1\}$, where the
  last component represents the only free remaining variable, i.e. $\bm X_2$. \\
The
  automaton $\mathcal A_\exists$ is built by starting from the one for $\varphi(\bm X_1,\bm X_2)$,
  and changing the transition labels from 
$(a,0,0)$ and $(a,1,0)$ to $(a,0)$;   
$(a,0,1)$ and $(a,1,1)$ to $(a,1)$, and analogously for those with $b$.   
The idea is that this last automaton nondeterministically ``guesses'' the quantified component
(i.e. $\bm X_1$) when reading its input, and the  resulting word $w \in (\Sigma
\times \{0,1\}^2)^*$ is such that $w \models \varphi(\bm X_1,\bm X_2)$. Thus,
$\mathcal A_\exists$ recognizes $\exists \bm X_1 (\varphi(\bm X_1, \bm
  X_2))$.
\end{itemize}

We refer the reader to the available literature for a full proof of equivalence
between the logic formula and the constructed automaton. Here we illustrate the
rationale of the above construction through the following example.

\begin{example}\label{ex:reg:log2aut}
 Consider the language $L = \{a,b\}^* a a \{a,b\}^*$: it consists of the strings satisfying
 the formula:\\
$
\varphi_L = \exists \bm x \exists \bm y (\sux(\bm x, \bm y) \land a(\bm x) \land
a(\bm y))$.

As seen before, first we translate this formula into a version using only
second-order variables:
$ \varphi'_L = \exists \bm X, \bm Y (
\Sing(\bm X) \land \Sing(\bm Y) \land \Sux(\bm X, \bm Y) \land 
\bm X \subseteq \bm W_a \land
\bm Y \subseteq \bm W_a)$.

The automata for $\Sing(\bm X)$ and $\Sing(\bm Y)$ are depicted in Figure \ref{fig:Sing-X-Y}; they could also be obtained by expanding the definition of $\Sing$ and
then projecting the quantified variables.
\begin{figure}
\[
\begin{array}{cc}
\begin{tikzpicture}[every edge/.style={draw,solid}, node distance=3.7cm, auto, 
                    every state/.style={draw=black!100,scale=0.5}, >=stealth]

\node[initial by arrow, initial text=,state] (q0) {{\huge $q_0$}};
\node[state] (q1) [right of=q0, accepting] {{\huge $q_1$}};

\path[->]
(q0) edge [loop above]  node {$(\circ,0,\circ)$} (q0)
(q0) edge [above] node {$(\circ,1,\circ)$} (q1)
(q1) edge [loop above]  node {$(\circ,0,\circ)$} (q1);
\end{tikzpicture}
&
\begin{tikzpicture}[every edge/.style={draw,solid}, node distance=3.7cm, auto, 
                    every state/.style={draw=black!100,scale=0.5}, >=stealth]

\node[initial by arrow, initial text=,state] (q0) {{\huge $q'_0$}};
\node[state] (q1) [right of=q0, accepting] {{\huge $q'_1$}};

\path[->]
(q0) edge [loop above]  node {$(\circ,\circ,0)$} (q0)
(q0) edge [above] node {$(\circ,\circ,1)$} (q1)
(q1) edge [loop above]  node {$(\circ,\circ,0)$} (q1);
\end{tikzpicture}
\end{array}
\]
 \caption{Automata for $\Sing(\bm X)$ and $\Sing(\bm Y)$.}
  \label{fig:Sing-X-Y}
\end{figure}
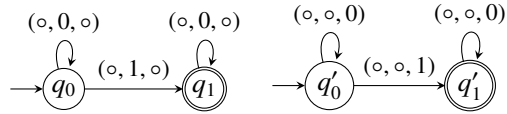

%
By intersecting the automata for $\Sing(\bm X)$, $\Sing(\bm Y)$, and $\Sux(\bm
X, \bm Y)$, by means of the customary construction of the cartesian product automaton (the details of the construction are not shown), we obtain an automaton that is identical to the one we defined for
translating formula $\Sux(\bm X_1, \bm X_2)$, where here $\bm X$ takes the role
of $\bm X_1$ and $\bm Y$ of $\bm X_2$.
Intersecting
it with those for $\bm X \subseteq W_a$ and $\bm Y \subseteq W_a$
produces the automaton of Figure \ref{fig:combination}.

\begin{figure}
\begin{center}
\begin{tikzpicture}[every edge/.style={draw,solid}, node distance=3.7cm, auto, 
                    every state/.style={draw=black!100,scale=0.5}, >=stealth]

\node[initial by arrow, initial text=,state] (q0) {{\huge $q''_0$}};
\node[state] (q1) [right of=q0] {{\huge $q''_1$}};
\node[state] (q2) [right of=q1, accepting] {{\huge $q''_2$}};

\path[->]
(q0) edge [loop above]  node {$(a,0,0)$} (q0)
(q0) edge [loop below]  node {$(b,0,0)$} (q0)
(q0) edge [above] node {$(a,1,0)$} (q1)
(q1) edge [above] node {$(a,0,1)$} (q2)
(q2) edge [loop above]  node {$(a,0,0)$} (q2)
(q2) edge [loop below]  node {$(b,0,0)$} (q2);
\end{tikzpicture}
\end{center}
\caption{Automaton for the conjunction of $\Sing(\bm X)$, $\Sing(\bm Y)$, $\Sux(\bm X, \bm Y)$, $\bm X \subseteq W_a$, $\bm Y \subseteq W_a$ .}
  \label{fig:combination}
\end{figure}
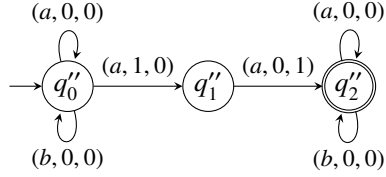

Finally, by projecting on the quantified variables $\bm X$ and $\bm Y$ we obtain
the automaton for $L$, given in Figure \ref{fig:automatonL}.
\qed 

\begin{figure}
\begin{center}
\begin{tikzpicture}[every edge/.style={draw,solid}, node distance=3.7cm, auto, 
                    every state/.style={draw=black!100,scale=0.5}, >=stealth]

\node[initial by arrow, initial text=,state] (q0) {{\huge $q''_0$}};
\node[state] (q1) [right of=q0] {{\huge $q''_1$}};
\node[state] (q2) [right of=q1, accepting] {{\huge $q''_2$}};

\path[->]
(q0) edge [loop above]  node {$a,b$} (q0)
(q0) edge [above] node {$a$} (q1)
(q1) edge [above] node {$a$} (q2)
(q2) edge [loop above]  node {$a,b$} (q2);
\end{tikzpicture}
\end{center}
\caption{Automaton for $L = \{a,b\}^* a a \{a,b\}^*$.}
  \label{fig:automatonL}
\end {figure}
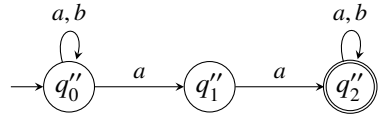

\end{example}

\section{Discussion}\label{sec:discussion}

The logical characterization of a class of languages, together with the decidability of the associated \textit{containment} problem (i.e., checking whether a language is a subset of another language in that class), is the main door towards automatic verification techniques. Suppose that a logic formalism $\mathfrak{L}$ is recursively equivalent to an automaton family $\mathfrak{A}$; then, one can use a formula $\varphi_\mathfrak{L}$ of $\mathfrak{L}$ to specify the requirements of a given  system and an abstract machine $\mathcal{A}$ in $\mathfrak{A}$ to implement the desired system: the correctness of the design defined by $\mathcal{A}$ w.r.t. to the requirements stated by $\varphi_\mathfrak{L}$ is therefore formalized as $L(\mathcal{A}) \subseteq L(\varphi_\mathfrak{L})$, i.e., all behaviors realized by the machine are also satisfying the requirements. This is just the case with FSAs and MSO logic for Regular Languages. 

Unfortunately, known theoretical lower bounds state that the decision of the above containment problem is PSPACE-complete and therefore intractable in general. The recent striking success of model-checking \cite{Emerson90}, however, has produced many refined results that explain how and when practical tools can produce results of ``acceptable complexity'' -- although the term ``acceptable'' is context-dependent, since in some cases even running times of the order of hours or weeks can be considered acceptable. In a nutshell, normally---and roughly---we trade a lower expressive power of the adopted logic, typically linear temporal logic, for a complexity that is ``only exponential'' in the size of the logic formulas, whereas the worst case complexity for MSO logic can be even a non-elementary function \cite {DBLP:FrickG04}.\footnote{There are, however, a few noticeable cases of tools that run satisfactorily at least in some particular cases of properties expressed in MSO logic \cite {MONA}.} In any case, our interest in these notes is not on the complexity issues, but it is focused on the equivalence between automata recognizers and MSO logics, which leads to the decidability of the above fundamental containment problem.